# Modulation of Superconductivity by Spin-canting in a Hybrid Antiferromagnet/Superconductor Oxide

Awadhesh Mani[a,b], T. Geetha Kumary[b], Daniel Hsu[a], J. G. Lin[a]*, and Chyh-Hong Chern[c]

[a]*Center for Condensed Matter Sciences, National Taiwan University, Taipei 106, TAIWAN*
[b]*Materials Science Division, Indira Gandhi Centre for Atomic Research, Kalpakkam 603102, INDIA*
[c] *Department of Physics, National Taiwan University, Taipei 106, TAIWAN*

The proximity effect of a C-type antiferromagnet (C-AFM) with the spin canting at low temperature is investigated in the hybrid $Nd_{0.35}Sr_{0.65}MnO_3$(NSCO)/$YBa_2Cu_3O_7$(YBCO) oxide system through magnetic and transport measurements. It is found that the onset of a spin-canted state destroys partially the superconducting order parameter. Interestingly, due to the instability of this spin-canted state, zero-resistivity recovers at the offset of spin canting. Our result demonstrates clearly the high sensitivity of superconducting order parameter to a modulation of internal field.

---

* Corresponding Author: Electronic mail: jglin@ntu.edu.tw

The studies on the proximity effects of the superconducting/magnetic (S/F) hybrid system started around 1980's, and continued with a longstanding importance due to the novelty of its underlying physics and the potential of its technological applications. Many exotic physical phenomena have been observed in the conventional S/F systems, such as the oscillation of superconducting transition temperature $T_c$ with respect to the thickness of F-layer, the formation of π junctions, and the domain wall superconductivity etc.[1-4] Nevertheless, the interplay of magnetism and superconductivity can be far more complex when it involves with the half metallic manganites coupled with the non-conventional high temperature superconductors (HTS).[5-15] Therefore, the full understanding of the magnetic proximity effect on HTS requires extensive studies on each individual hybrid system with various composition. In this work, the C-type antiferromagnetic (AFM) $Nd_{0.35}Sr_{0.65}MnO_3$ (NSMO) is chosen to epitaxially grow on the top of an $YBa_2Cu_3O_7$ (YBCO) film. The C-type AFM state is characterized by an apically elongated $MnO_6$ octahedral, resulting from the ordering of the rod-type $d(3z^2-r^2)$ orbitals.[16] Its Neel temperature[17] $T_N$ is ~270 K, below which spins are aligned ferromagnetically along *c*-direction but antiferromagnetically coupled within the *a-b* plane. It is also believed that there exists a phase transition to the canted AFM state at $T_{CAF}$ = 45K, although the ferromagnetic signal is extremely faint in the neutron scattering experiment. In this work we apply the hybrid system of NSMO/YBCO to probe the magnetic nature of NSMO. It is prevailed that the internal magnetic field induces the vortex dissipation in superconducting state and generates a finite resistance. More importantly, the phase transition in NSMO/YBCO can be shown unambiguously to be of the first order type.

Hybrid system of NSMO(40nm)/YBCO(70nm), denoted as NY47, is deposited on (*100*) $LaAlO_3$ (LAO) single crystal substrate with a commercial pulsed laser deposition system (Neocera Pioneer 180). Films of pure YBCO, NSMO are also grown for comparison. During the film-growth, the substrate temperatures are kept at 850 and 800 °C in the flowing $O_2$ atmosphere of 50 mTorr for YBCO and NSMO films, respectively. The hybrid sample is further post annealed at 400 °C in-situ for 60 min under an oxygen pressure of 300 Torr. The crystal structures of all films are analyzed by X-ray diffraction (XRD). The temperature ($T$) dependent resistivity ($\rho$) is measured using the standard four-probe method for the films with the dimension of 10×2.5 mm$^2$. Magnetizations ($M$) are measured as functions of $T$ and applied field ($H$) by a superconducting quantum interference device system (Quantum Design).

The XRD patterns of NSMO, YBCO and NY47 are depicted in Fig. 1. The presence of only *(00l)* lines suggests that all the films are grown with *c*-axis perpendicular to the substrate surface. Furthermore, the lines in NY47 match well with those of NSMO and YBCO, indicating its phase purity and the formation of requisite structure. The image of high resolution transmission electron microscope confirms the epitaxial growth of bilayer sample (not shown here). Based on the positions of *(00l)* lines, the *c*- lattice parameters of YBCO and NSMO phases can be separately calculated. Accordingly, the *c*-parameter of YBCO-phase is 11.691 Å, which is comparable with the value of the bulk YBCO (*c*=11.6701 Å, *a*=3.8820 Å, *b*=3.8148 Å; Orthorhombic structure; space group: Pmmm). On the other hand, the *c*-parameter of NSMO-phase in NY47 is found to be 7.809 Å, which is larger than that of the bulk NSMO ( *c* = 7.7748 Å, *a* = b = 5.3670 Å; Pseudo tetragonal Phase; space group: I4/mcm), implying that the tensile strain exists in the top NSMO layer of NY47.

The curves of M(T) for the NSMO, NY47 and YBCO samples with the ZFC mode under 200 Oe parallel to the film-surface are shown in Fig. 2. It is found that $T_N$ is ~ 270 K for both NSMO and NY47 samples. The superconducting transition temperature $T_c$ is ~ 88 K for YBCO and ~ 83 K for NY47. The marginal suppression of $T_c$ by AFM interaction could be attributed to the nesting of Fermi surfaces of AFM-band associated with the symmetry-breaking in the momentum space.[18] At lower T, another magnetic transition from C-type AFM to a canted AFM state is seen in both NSMO and NY47 at different transition temperatures $T_{CAF}$. The onset of $T_{CAF}$ is ~ 44 K in NSMO, but 34 K in NY47 as indicated by arrows. The *M-H* curves of NSMO at 10 and 40 K are plotted in the inset of Fig. 2(a), which shows a hysteresis loop with saturation magnetization of ~10 emu/cm$^3$ at 40 K (see the solid lines) and a linear line (see the dashed one) without hysteresis at 10 K. In the thin film form of the NSMO, the magnetic transition looks like the second-order type, but in NY47 it looks to be the first-order type. The distinct difference in the nature of phase-transition between NSMO and NY47 could be due to the dimensionality since the thick NY47 film (110 nm) can be treated as a bulk. It will be shown later that the transport data also suggests that the spin-canting transition is of the first-order type in NY47. Furthermore, the sudden drops of M at T < 18 K for NSMO and at T < 24 for NY47 imply a switch-off of spin-canting. Inset of Fig. 2(b) shows the amplified M(T) curve of NY47 near the canting state. Two different ordering states of NY47 below $T_{CAF}$ can be clearly seen in the *M-H* curves at 30 and 5 K as shown in Figs. 3(a) and 3(b) respectively. At 30 K, the shape of *M-H* curve resembles a FM system, but with a marked asymmetry along horizontal (*H*) as well as vertical (*M*) axes. According to the literature, the asymmetry with respect to *H* is due to the interaction between the eddy current and the magnetic moment during the field cooled process; while the asymmetry with respect to *M* is the effect of magnetic pinning.[19] At 5 K, the *M-H* curve corresponds to a typical superconducting state of star-like shape.

Figures 4(a) and (b) display the ρ(T) of NY47 at various $H$ and NSMO at zero field respectively. The positions of Neel temperatures in NSMO and NY47 are marked by arrows in 4(b) and the upper inset of 4(a) respectively. NSMO behaves as an insulator and its ρ(T) exhibits a dip near $T_{CAF}$; while NY47 possesses zero-resistance around 60 K and it becomes resistive again at $T_{CAF}$. The field dependency of ρ(T) in NY47 shows a typical shift toward to low temperatures, which is attributed to the vortex-dissipation due to the internal magnetic field from the magnetic moments of the spin-canting state. The discontinuity of ρ at $T_{CAF}$ implies that the transition is of the first-order type, where

$$\rho/\rho_n = B/H_{c2}$$

with $\rho_n$ being the resistance of the normal state of YBCO, $H_{c2}$ the upper critical field, and B proportional to the magnetic moment of the NSMO in the NY47 sample. The switching-off of spin-canting state at 24K shows complicate resistive transitions because of the competition of the external magnetic field and the internal magnetic field.

The distinct changes of M(T) and ρ(T) near $T_{CAF}$ and $T_{re}$ are also reflected in their magnetoresistance (MR) behavior as shown in the lower inset of Fig. 4(a), with MR-value defined as [ρ(1tesla)- ρ(0)]/ρ(0). The MR-value is negative with a maximum of - 46 % near $T_{CAF}$, and it switches to a positive value of ~ 60% around $T_{re}$. The negative MR in the normal state region is consistent with the picture that the magnetic field enhances the spin ordering and hence reduces the spin scattering contribution to the resistivity. As to the positive MR-value in the superconducting regime, it could be due to a combination of many factors, including[20,21]: (1) the Aslamazov-Larkin (AL) term arising from the fluctuation of Cooper pairs, (2) the regular and anomalous Maki-Thompson (MT) contribution originating from the coherent scattering of a Cooper pair, (3) the correction of density of states (DOS) in association with the forming of bound fluctuated pairs, and (4) the vortex dissipation.

In conclusion, we successfully probe the nature of the spin-canting transition and the related magnetic phase using the hybrid $Nd_{0.35}Sr_{0.65}MnO_3$/$YBa_2Cu_3O_7$ system. In superconducting state the internal magnetic field induces the proliferation of the vortices, leading to the resistive motion by applying current. The modulation of the superconductivity by a small internal field instead of a large external field may be used to study other magnetic systems, particularly for the ones with their magnetic phase transitions below the superconducting transition temperature of $YBa_2Cu_3O_7$.

ACKNOWLEDGMENTS

This work is partly supported by the National Science Counsel of R. O. C. (Grant Nos. NSC-96-2112-M-002-027-MY3 and NSC-97-2112-M-002-027-MY3). C.-H. Chern is also supported by the Project No. 97R0034-25 from the National Taiwan University.

**Figure Captions**

**Fig. 1.** The XRD patterns of NSMO, YBCO and NY47. Note that all the peaks are identified as *(00l)*.

**Fig. 2.** (a)-(c)**:** Zero-field-cooled (ZFC) magnetization (*M*) vs temperature (T) for NSMO, NY47 and YBCO respectively. The dash line marks $T_N$ and arrows indicate $T_{CAF}$ and $T_c$. Inset in (a) is the amplifies M(T) near the apin-canting transition and Inset in (b) represents the *M-H* curves of NSMO at 40 and 10 K with respect to the solid lines and dashed line.

**Fig. 3.** *M-H* curves of NY47 samples at (a) 30 K and (b) 5 K.

**Fig. 4.** (a) and (b) plot the temperature dependent resistivity, ρ(T), of YN47and NSMO respectively. The lower inset in Fig. 4(a) displays the temperature dependent MR measured at a moderate field of 1Tesla, while the upper inset shows the expanded ρ(T) of NY47 near the Neel temperature.

**Fig. 1.** Mani et al.

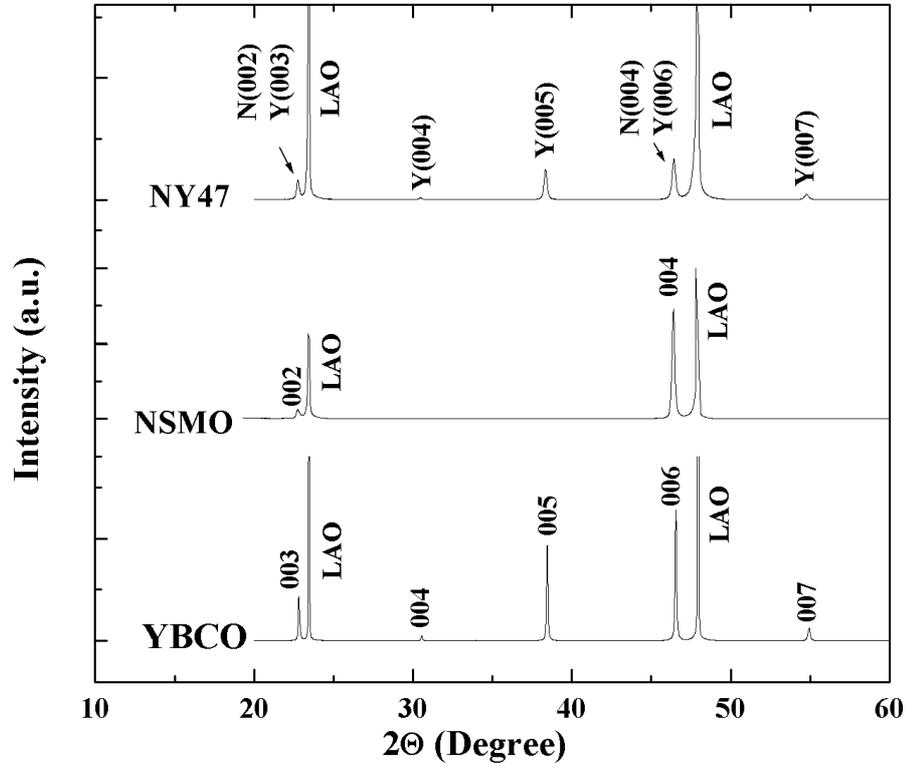

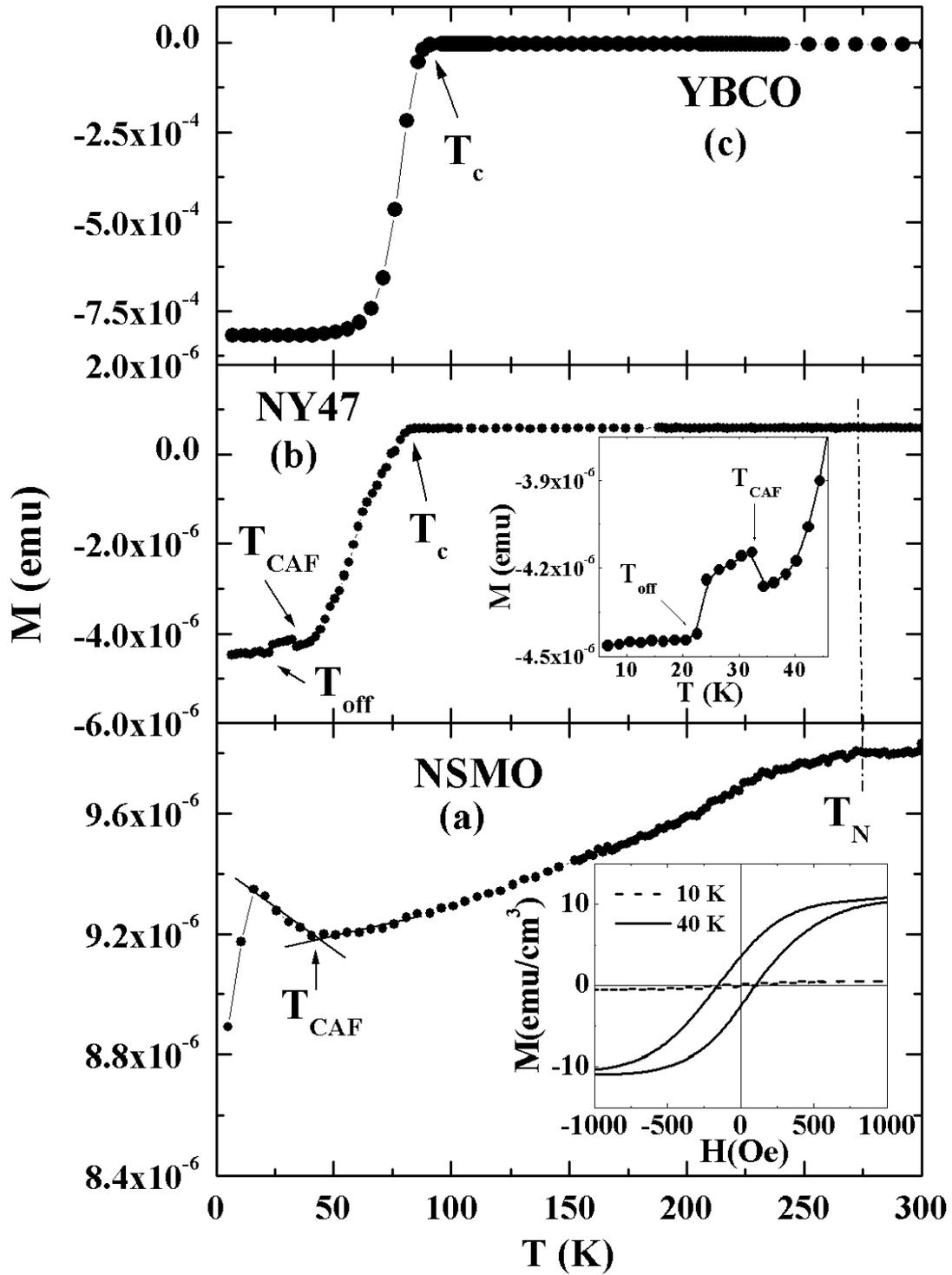

**Fig. 3.** Mani et al.

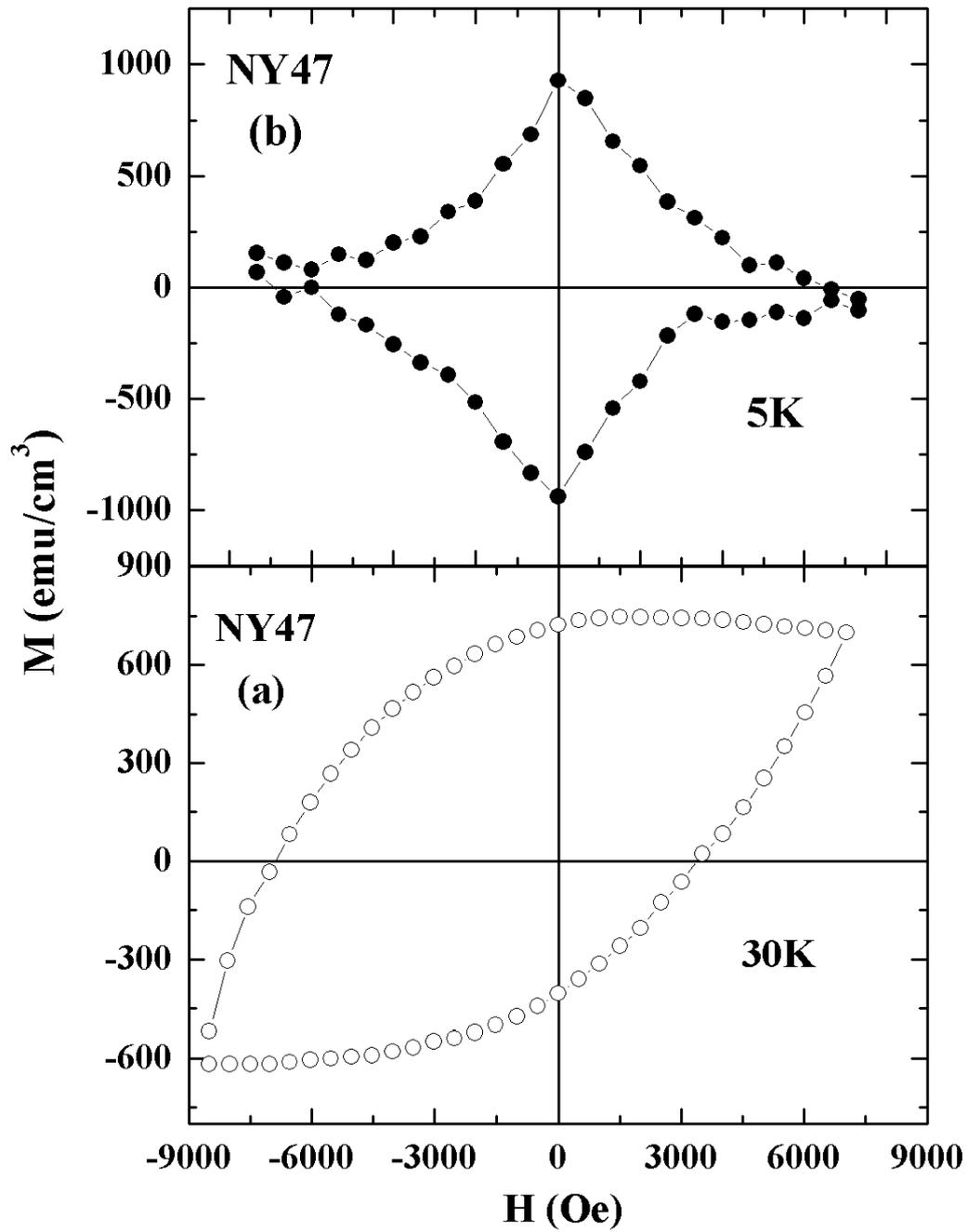

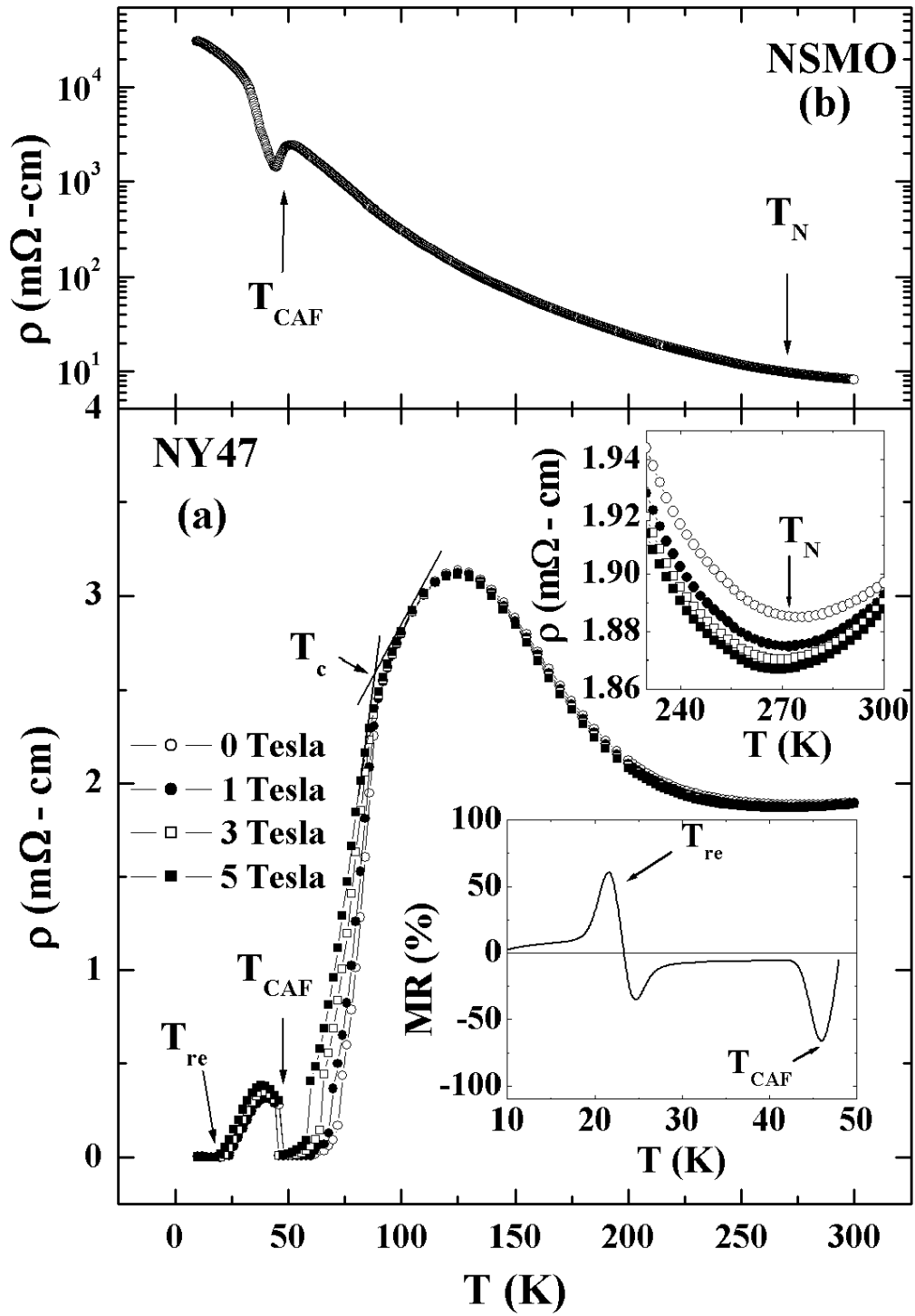